\documentclass{article}
\usepackage{spconf,amsmath,graphicx,xcolor,subcaption,array,tikz,hyperref}
\usepackage[ruled,vlined]{algorithm2e}

\newcommand\copyrighttext{%
  \footnotesize Copyright 2020 IEEE. Published in the IEEE 2020 International Conference on Image Processing (ICIP 2020), scheduled for 25-28 October 2020 in Abu Dhabi, United Arab Emirates. Personal use of this material is permitted. However, permission to reprint/republish this material for advertising or promotional purposes or for creating new collective works for resale or redistribution to servers or lists, or to reuse any copyrighted component of this work in other works, must be obtained from the IEEE. Contact: Manager, Copyrights and Permissions / IEEE Service Center / 445 Hoes Lane / P.O. Box 1331 / Piscataway, NJ 08855-1331, USA. Telephone: + Intl. 908-562-3966.}
\newcommand\copyrightnoticeA{%
\begin{tikzpicture}[remember picture,overlay]
\node[anchor=south,yshift=10pt] at (current page.south) {\fbox{\parbox{\dimexpr\textwidth-\fboxsep-\fboxrule\relax}{\copyrighttext}}};
\end{tikzpicture}%
}

\title{Learning to Model and Calibrate Optics via\\  a Differentiable Wave Optics   Simulator}
\name{Josue Page$^{\star \dagger}$, Paolo Favaro$^{\star}$ \thanks{J. Page and P. Favaro acknowledge the support of the UniBe ID Grant ``A deep learning approach to light field microscopy for volume imaging in life sciences.''}}
\address{$^{\star}$ University of Bern, Switzerland \\
      $^{\dagger}$ Technical University of Munich}
\begin{document}
%
\maketitle
\copyrightnoticeA
\begin{abstract}
We present a novel learning-based method to build a differentiable computational model of a real fluorescence microscope. Our model can be used to calibrate a real optical setup directly from data samples and to engineer point spread functions by specifying the desired input-output data.
This approach is poised to drastically improve the design of microscopes, because the parameters of current models of optical setups cannot be easily fit to real data. 
Inspired by the recent progress in deep learning, our solution is to build a differentiable wave optics simulator as a composition of trainable modules, each computing light wave-front (WF) propagation due to a specific optical element. 
We call our differentiable modules WaveBlocks and show reconstruction results in the case of lenses, wave propagation in air, camera sensors and diffractive elements (e.g., phase-masks).
\end{abstract}
\section{Introduction}
\label{sec:intro}

\begin{figure}[!]
    \centering
    \includegraphics[width=0.45\textwidth]{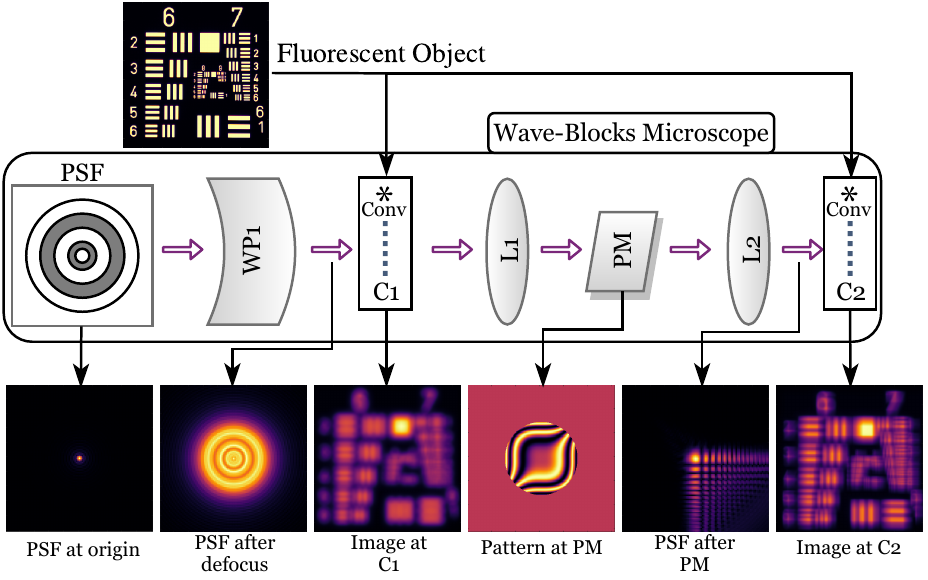}%
    \caption{Fluorescent microscope recreated with WaveBlocks. A bright-field PSF is propagated through air by WP1, then imaged by the first camera (C1). Alternatively, the wave-front continues to the 4-f system (L1-2) with a phase-mask (PM) placed at its Fourier plane, later  the second Camera (C2) convolves the object and the PSF from the back focal plane of L2. Each camera (C1-C2) is used for a separate experiment in sec.~\ref{sec:experiments}.}
    \label{fig:WB_microscope}
\end{figure}

Microscopes have a fundamental impact in biology, life science and engineering. Because of their essential role in imaging, their design has seen a constant evolution in an effort to improve the imaging quality (e.g., the lateral and axial resolutions and the speed). 
Towards this goal, one approach is to improve the quality of the captured data through better sensors, optics and their configuration; also referred to as \emph{point spread function (PSF) engineering} \cite{Klauss2017,Shuang2016,Zhao2013}, and it is currently achieved mostly through manual design and analysis.
A second approach is instead to improve the captured data via image processing. The task is posed as an inverse problem, where the latent high-quality image is reconstructed by combining a generative model of observed data (the data fidelity term) with a model of the patterns typically observed in the data (the data prior).
In this domain recent deep learning methods have demonstrated an impressive performance, both in terms of image reconstruction accuracy and speed, thanks to their ability to efficiently and accurately describe complex data priors and generative models \cite{Wu2019}.

Current PSF engineering is sub-optimal as it does not take into account the subsequent image processing step. In fact, an optimal data capture system should be designed around the patterns of typically observed data, and this is an extremely difficult problem to solve analytically.
Therefore, we argue that a better approach is to jointly optimize both steps numerically through real data.
Towards this goal, in this paper we introduce and study a model of optics that can be readily integrated with deep learning models for image processing. 
Our solution is to first accurately describe light propagation through optical elements (e.g., lenses, air, phase-masks) with  computational modules, which we call \emph{WaveBlocks} (see Fig.~\ref{fig:WB_microscope}). In this way, it is possible to relate each module to a physical counterpart. Secondly, to enable the automatic optimization of the optical elements, we ensure that the modules are differentiable with respect to their parameters.
We can then fit real data to our compositional model through optimization (i.e., via stochastic gradient descent), a procedure that we call \emph{calibration}. 
In this paper, we illustrate our approach with several calibration tasks. First, we show how to accurately fit the PSF of a real microscope by using real data. Later, we show how to recover the diffracting pattern caused by the lack of 100\% fill factor in a spatial light modulator (SLM) within a 4-f setup.

\section{Previous Work}
PSF engineering depends on the imaging properties of interest, for example, one could aim to improve imaging for super-resolution \cite{Nehme2019, Durand2018, Shuang2016,Grover2012,Klauss2017}, or 3D reconstruction \cite{Cohen2014,Wu2019,Shuang2016, Quirin2013}, or extended depth of field \cite{Zhao2010,Quirin2013,Zhao2013}, or even light-field microscopy \cite{Cohen2014}. 
General phase-mask optimization \cite{Zhao2013} and calibration \cite{Klauss2017} have been explored before. One approach to solve calibration is to pose the task as a blind deconvolution problem \cite{Soulez2012a,Lim2019,Kim2015}. Also phase retrieval and the removal of the zero-order diffraction from a SLM have been previously studied \cite{Zhang2009,Liang2012}. 
Previously a microscope simulator for the generation of synthetic biological images was proposed \cite{Rajaram2012}. However, this computational model is not suitable for data-driven automatic optimization of the optical components.
Thus, our proposed WaveBlocks, a differentiable and scalable optimization framework, provides a fundamental tool to do automatic data-driven PSF engineering.
In future, work we aim to demonstrate our framework on a variety of image processing tasks and imaging devices.

\section{Building a Wave Optics Simulator}

As discussed in the Introduction, we are interested in building a differentiable model that can accurately approximate real optics and with modules that are associated to each optical component in a microscope. We start by defining the image formation process of a fluorescence microscope. In its simplest instance, it follows the convolution operation
\begin{equation}
    i = H \ast o + n,
    \label{eq:ImgForm}
\end{equation}
where $o$ is the fluorescent object, $i$ the observed image, $n$ additive noise and $H$ the PSF or system matrix. $H$ can be acquired in several ways: by measuring it directly with the microscope, or, as it is usually done, by modeling it computationally. In our modules, the light emitted by a fluorescent particle is propagated as a complex WF through each of the optical elements in the microscope until it hits the sensor, where the irradiance of the WF is computed and stored as the PSF for later use.

The linear operators that model each optical element forming the PSF are based on complex diffraction integrals, which difficult to optimize when stacked together. However, their linearity enables the usage of each operator as a block that receives an input WF and produces an output. Moreover, it is possible to easily stack as many blocks as desired. We implement these operators in the Pytorch framework \cite{pytorch}, which performs auto-differentiation. The architecture is user-friendly and allows the optimization of a wide variety of parameters in an optical system. The rationale behind WaveBlocks is that the user provides a bright-field PSF (for example, generated in Fiji \cite{Schneider2012}). This PSF propagates through the user-defined optical blocks until it reaches a camera module. On this module the image irradiance is computed and convolved with a user-provided object as shown in Fig.~\ref{fig:WB_microscope}.

\subsection{Wave Blocks Implementation}
In the following sections, we describe the computational models of the key optical elements implemented so far.

\subsubsection{Wave propagation (WP)}
To compute the monochromatic WF propagation through a medium (the WP block), we use the Rayleight-Sommerfeld integral (see \cite{Voelz2011}, page 52). 
Our implementation of the Rayleight-Sommerfeld propagation is limited to a minimum distance of $200\mu m$, to make sure that the required sampling remains computationally practical.
According to the Fourier convolution theorem (see  \cite{Voelz2011}, page 39), the WP block can be obtained via
    $U_2(x,y) = \boldsymbol{F}^{-1}\{\boldsymbol{F}\{U_1(x,y)\}\boldsymbol{F}\{h(x,y)\}\}, \label{eq:FCT}$ with 
    $h(x,y) = \frac{z}{j\lambda} \frac{\exp(jkr)}{r^2} \nonumber$, 
where $\boldsymbol{F}$ denotes the Fourier transform, $\boldsymbol{F}^{-1}$ denotes its inverse, $(x,y,z)$ are 3D image spatial coordinates, $j$ denotes the imaginary component in complex numbers, $h(x,y)$ is the Rayleigh-Sommerfeld impulse response, $\lambda$ is the light wavelength, $k$ is the wave number and $r=\sqrt{z^2+x^2+y^2}$ is the distance from a point in $U_1$ to another in $U_2$.

\subsubsection{Lens (L)}
The lens block (the L block) describes a convex lens which propagates a WF from a plane at the front of the lens to the back of it. 
According to the Fraunhofer approximation (see \cite{Voelz2011}, page 96), the lens propagation from the front plane to the back focal plane is equivalent to a scaled Fourier transform given by the following equation
    $U_2(x,y) = c(x,y) \cdot \boldsymbol{F}\{U_1(x,y)\cdot P(x,y)\} \cdot dx_1^2$, 
where $d x_1$ is the source sample interval, $f_l$ the focal length of the lens, $P(x,y)$ the pupil function and $c$ is the scaling factor given by
$c(x,y) =  \frac{\exp(jkf_l)}{j\lambda f_l} \exp\left[j\frac{k}{2f_l}(x^2+y^2)\right]$. 

\subsubsection{Camera (C)} \label{sec:camera}

The camera block (the C block) performs two tasks:
    1) It determines the PSF $H$ by computing the irradiance of the incoming WF, which is the time-averaged square magnitude of the field $U$, given by
    $H(x,y)=U(x,y)U(x,y)^* = |U(x,y)|^2.$
    This time averaging occurs due to the incapability of current detectors to follow the high frequency oscillations of the electric field ($>10^{14}Hz$) (see \cite{Voelz2011}, page 49); 
    2) It convolves the computed PSF $H$ with an object $o$ placed in front of the microscope, i.e.,
        $i = H\ast o.$

\subsubsection{Phase-mask (PM) or spatial light modulator (SLM)}
The phase-mask block (the PM block) describes a phase-mask that distorts the phase of the WF in a space-variant way. The incoming WF $U_1$ is modified by the modulation function $\phi$ as in 
 $U_2(x,y) = U_1(x,y) \exp(j\phi(x,y)).$
A special case applies when a phase-only SLM is used where only the imaginary part of the exponential is multiplied by $U_1$ and the real part becomes zero. In WaveBlocks $\phi$ is a parameter that can be changed at will or optimized simply by naming it in the initializer of the Microscope class.

\subsection{Data-driven calibration}

To calibrate an optical system we capture images of several known objects. If we denote with $H_{gt}$ the real system matrix, we can capture real images $i$ of known objects $o$, which we model as $i = H_{gt}\ast o$. WaveBlocks allow us to model this data via a system matrix $H$ parametrized via $\Theta$, which collects all the settings of the optical elements (e.g., the main objective PSF, the phase-mask phase change).
This optimization problem can be then written as the minimization of the expectation over all objects and corresponding observed images (which can be approximated by using a finite, albeit large, number of samples)
\begin{equation}
    \hat{\Theta} = \arg\min_\Theta \text{E}_{i,o}\left[\ell \left(o \ast H(\Theta),i \right) \right]
\end{equation}
where $\ell$ is the cost function measuring the discrepancy between the observed images $i$ and the synthesized ones $H(\Theta)\ast o$.
$H$ can be formed by the successive stack of blocks. For example, the configuration shown in Fig.~\ref{fig:WB_microscope} yields $H(\Theta)=C2(L2(PM(L1(WP1(PSF)))))$, where
$\Theta$ has the parameters of the PM and/or the PSF. This configuration is also used in the experiments in sec.~\ref{sec:sub:PM_estimation}.
\section{Experiments} \label{sec:experiments}
To demonstrate the capabilities of our proposed calibration approach, we aim to reconstruct two sets of parameters: 1) the initial PSF of the microscope and 2) the distortion diffraction pattern generated by the empty space between the pixels of a SLM.
We employ a USAF 1951 resolution target, a 20x 0.45NA objective, a 165mm focal length tube-lens, a camera (C1-C2) with 6.9$\mu m$ pixel size, two lenses (L1-L2) with focal length of 150mm and aperture of 50.8mm and the phase-mask (PM) used is an SLM Holoeye Pluto-vision.  

\subsection{Bright-field PSF estimation} \label{sec:sub:PSF_estimation}
In this section we focus on reconstructing the PSF of a real microscope in the simplest case, where only a PSF block representing the PSF generated by the objective and tube-lens, a wave-propagation block and a camera block are used (PSF, WP1 and C1 in Fig.~\ref{fig:WB_microscope}). As $o$ we choose the USAF 1951 target object (only one image in this case), and the loss is the normalized mean square error (NMSE) loss given by $\text{NMSE}(i,k)=\frac{||i-k||^2_2}{||i||_2^2}$. The optimization uses the Adam optimizer.

\subsubsection{Recovery of the PSF from synthetic data}
To measure the robustness of the PSF reconstruction algorithm we randomly transform and reconstruct a PSF computed via the scalar Debye theory \cite{BornWolf:1999:Book,Broxton2013}. First, we translate the object randomly in a range of $\pm$20 pixels, then re-scale it in $x$ and $y$ directions independently in the range 1$\pm$0.3 and defocus it using WP1 by a random distance between $\pm$200 and $\pm$1000 $\mu m$ in image space. This synthetic PSF $H_{gt}$ is built by setting the parameters to some $\Theta_{gt}$. A total of 500 different PSFs were estimated, using defocused images at -50, 0 and 50 $\mu m$ in reference of the focal plane of the objective, this by using the Adam optimizer from Pytorch. 
The results of this evaluation (see Table~\ref{table:synthResults}) show that even though the error from the PSF estimation is not optimal, the error of from the images used for optimization is very low, meaning that the system finds a PSF that its able to reproduce the observed images.

\begin{table}[]
\begin{tabular}{l|l|l|l|l|}
\cline{2-5}
& \multicolumn{2}{c|}{\textbf{$NMSE(i_{gt},i)$}} & \multicolumn{2}{c|}{\textbf{$NMSE(\Theta_{gt},\Theta)$}} \\ 
\cline{2-5} & \multicolumn{1}{c|}{\textbf{NMSE}} & \multicolumn{1}{c|}{\textbf{Std-Dev}} & \multicolumn{1}{c|}{\textbf{NMSE}} & \multicolumn{1}{c|}{\textbf{Std-Dev}} \\ 
\hline
\multicolumn{1}{|c|}{\textbf{\begin{tabular}[c]{@{}c@{}}PSF\end{tabular}}} & $9.34\cdot 10^{-3}$ & $6.95\cdot 10^{-3}$ & 1.5 & 0.43 \\ 
\hline
\multicolumn{1}{|c|}{\textbf{\begin{tabular}[c]{@{}c@{}}PM\end{tabular}}}   & $6.30\cdot 10^{-2}$ & $2.14\cdot 10^{-3}$ & 0.51 & 0.42 \\ 
\hline
\end{tabular}
\caption{Mean and standard deviation of NMSE error between the image stack ($i$) captured by $C1$ and $C2$ and the GT stack ($i_{gt}$). Also NMSE of the estimated parameters, PSF in experiment one, and PM in experiment two.}
\label{table:synthResults}
\end{table}

\subsubsection{Recovery of the PSF from real data}
In this experiment we aim to recover the PSF of a real microscope. We acquire a stack of images of the USAF 1951 target placed at depths spanning -50 to 50 $\mu m$ in steps of 2$\mu m$ relative to the focal plane of the objective. For this experiment only the images at depths -50, 50 and 0 $\mu m$ were used in the training, and the rest were stored for later testing. The recovered PSF can be seen in Fig.~\ref{fig:PSFs} next to the PSF of an ideal microscope. Notice how the recovered PSF exhibits aberrations not present in the ideal case.

\newcommand{\photo}[1]{%
    \includegraphics[width=1.5cm]{#1}
}
\setlength{\tabcolsep}{3pt}
\begin{figure}
    \centering
    \begin{tabular}{cccc}
    \multicolumn{4}{c}{\textbf{PSF recovery experiment sec. \ref{sec:sub:PSF_estimation}}} \\
\hline
Initial PSF & Image at C1 & GT Image & NMSE: 0.633\\ 
\hline
\photo{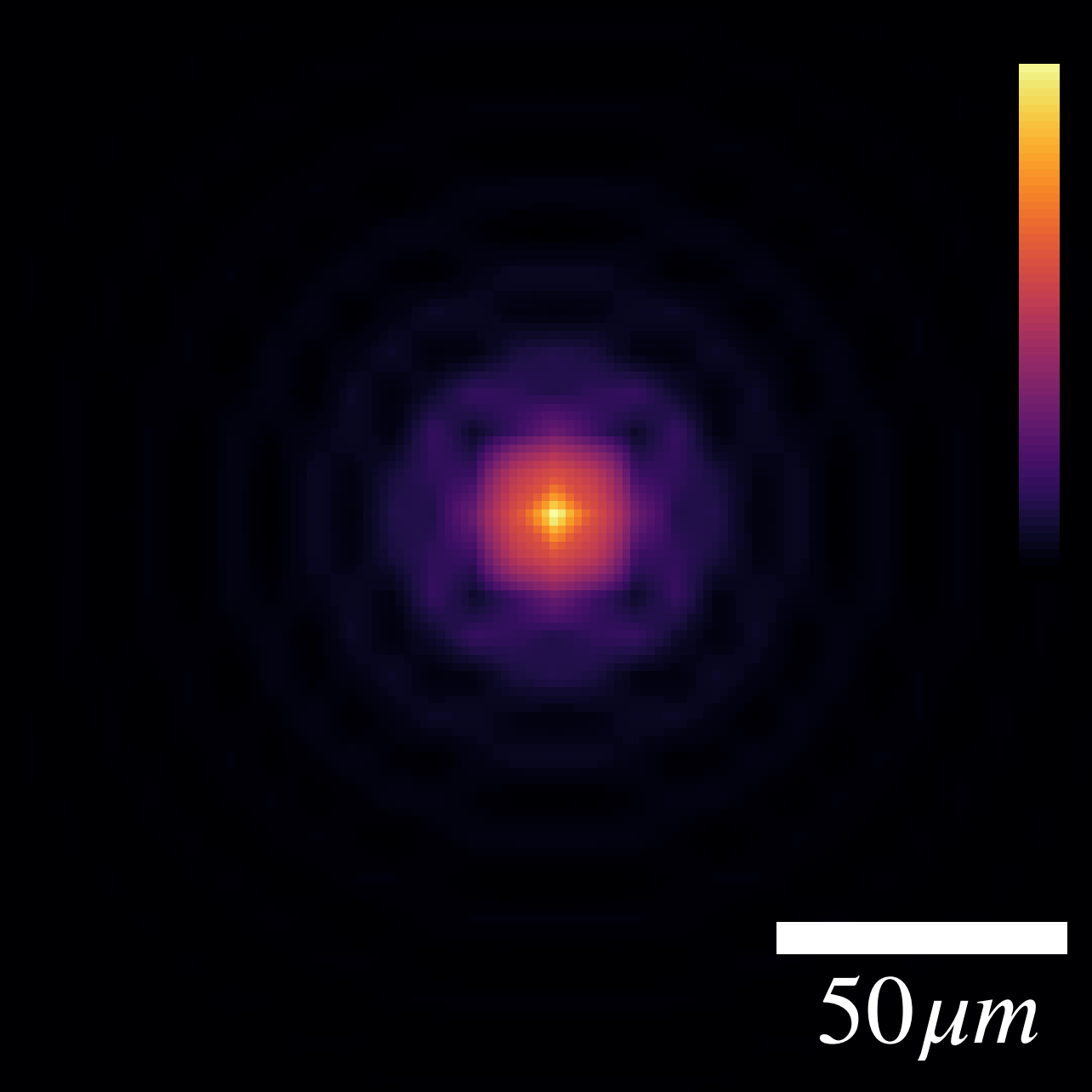} &  \photo{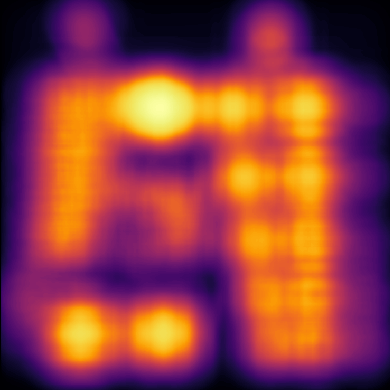} & \photo{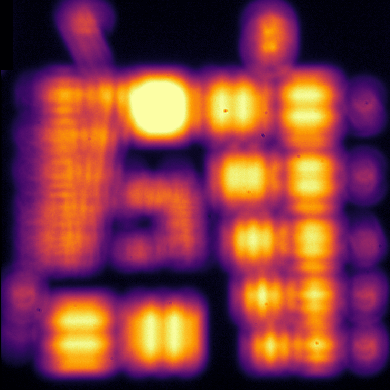} &  \photo{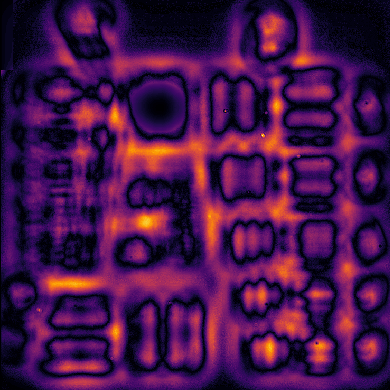} \\
\hline
Recovered PSF & Image at C1 & GT Image & NMSE: 0.520\\ 
\hline
\photo{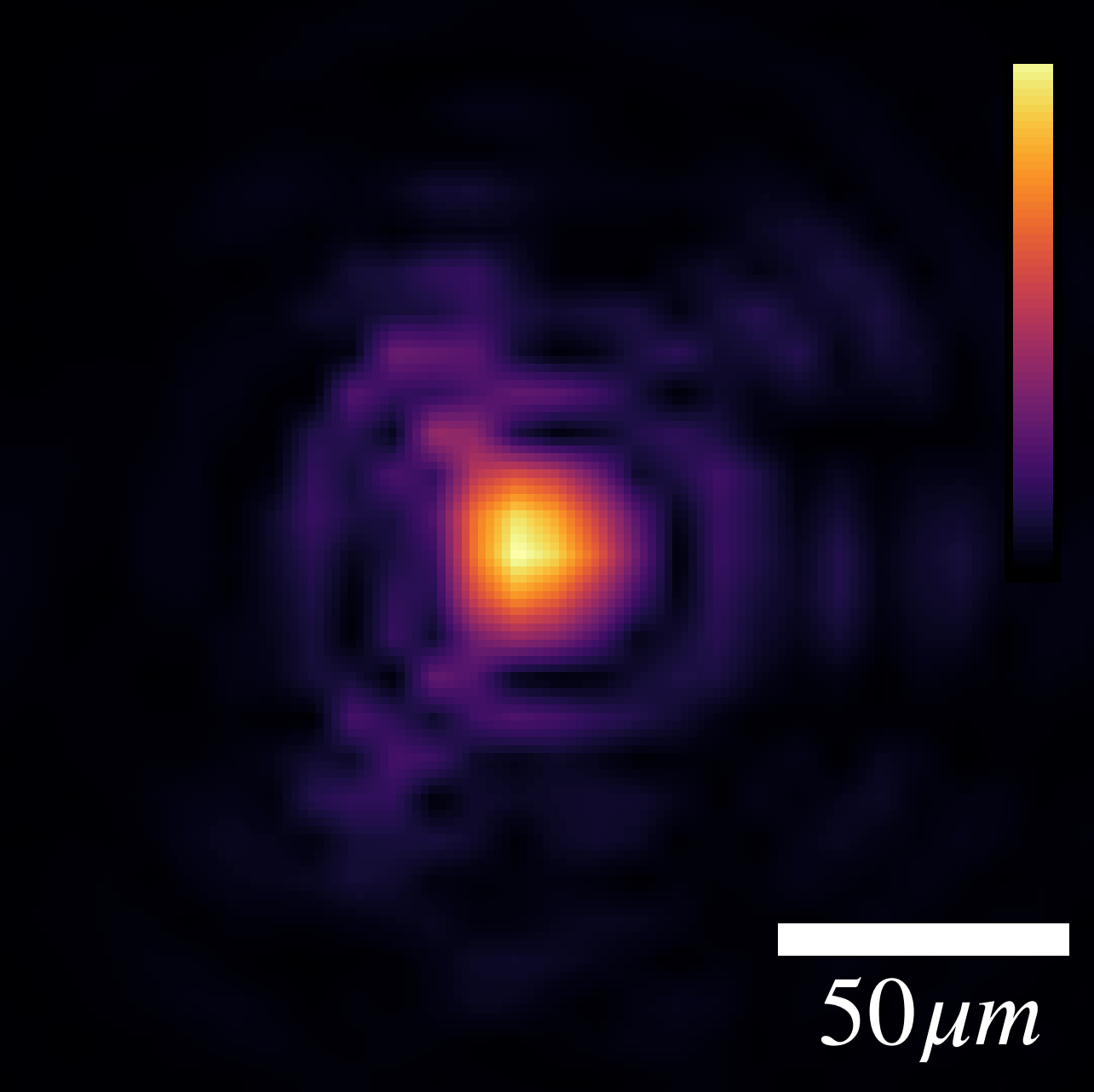} &  \photo{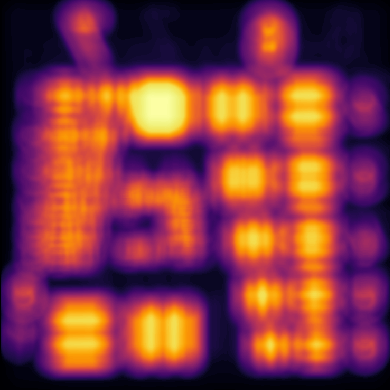} & \photo{figures/PSF_PM/PSF_GT_1.png} &  \photo{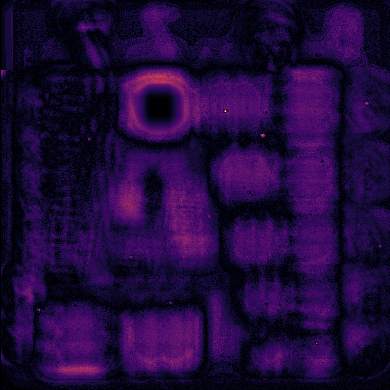} \\
\hline
\multicolumn{4}{c}{\textbf{Phase-Mask recovery experiment sec. \ref{sec:sub:PM_estimation}}} \\
\hline
Constant PM & Image at C2  & GT Image & NMSE: 0.153\\  
\hline
\photo{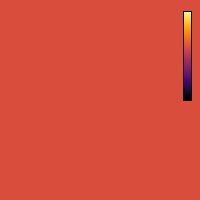} &  \photo{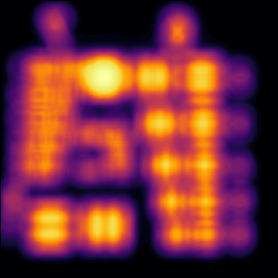} &  \photo{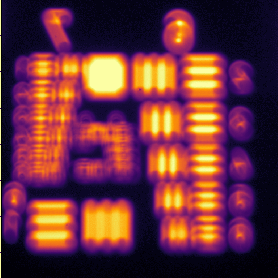} &  \photo{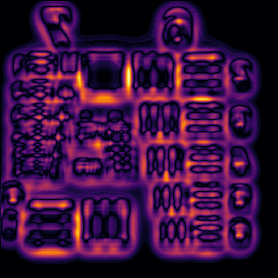} \\
\hline
Recovered PM & Image at C2  & GT Image & NMSE: 0.089\\  
\hline
\photo{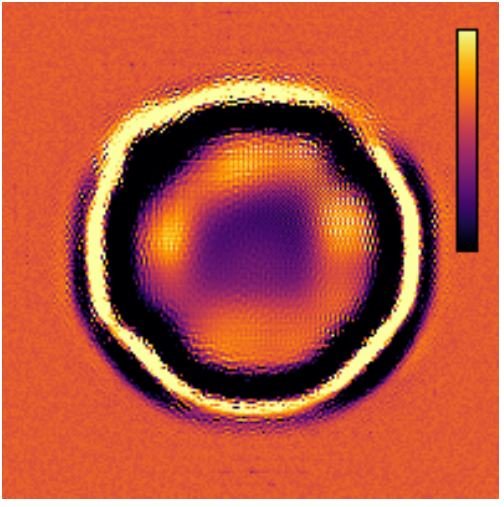} &  \photo{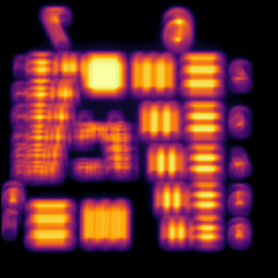} &  \photo{figures/PSF_PM/GT_PM_1.png} &  \photo{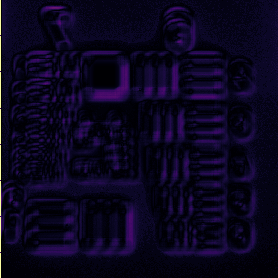} \\
\hline
\end{tabular}
    \caption{First row: comparison between the PSF (at $0\mu m$) of an ideal microscope and the recovered PSF obtained through our approach. Second row: comparison between a constant PM pattern and the recovered one.}
    \label{fig:PSFs}
\end{figure}





\begin{figure}[t]
    \centering
    \includegraphics[width=0.45\textwidth]{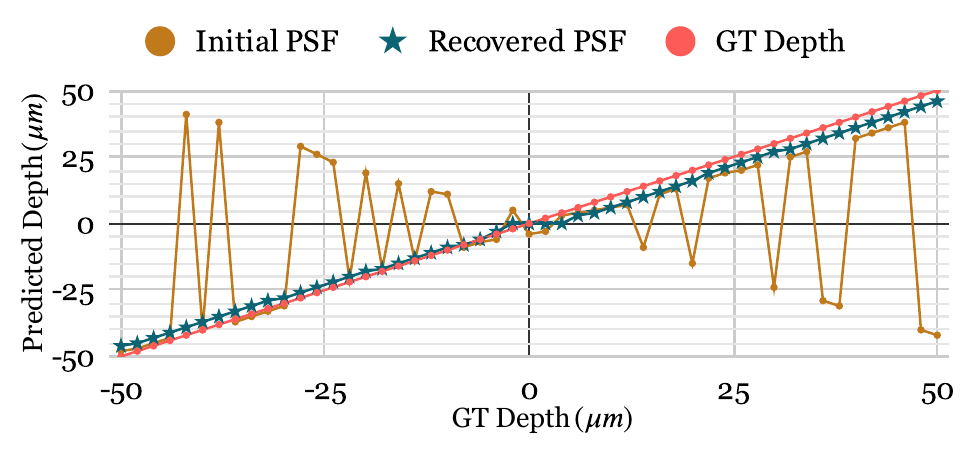}
        \includegraphics[width=0.45\textwidth]{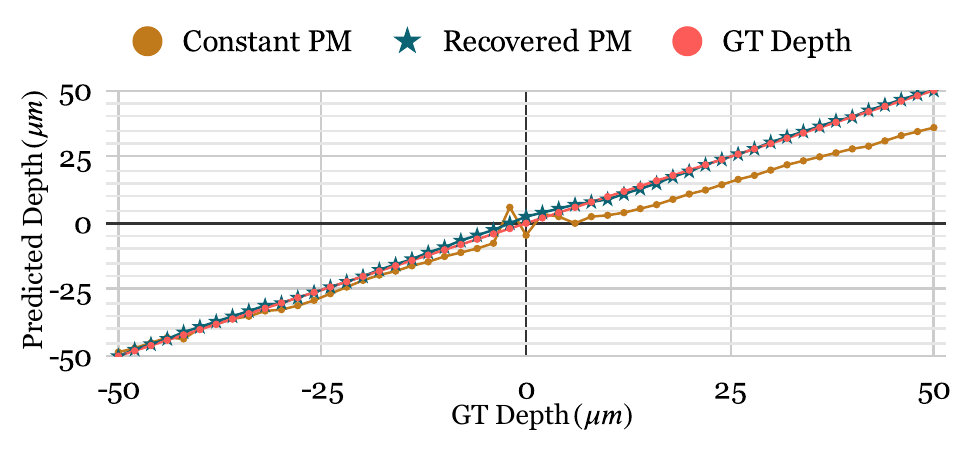}
    \caption{Top: Comparison of depth prediction using the initial PSF or the optimized one.
    Bottom: Comparison of depth prediction using the recovered PSF and either a constant PM as displayed in the SLM or the recovered PM.    \label{fig:PSF_depths}
    \label{fig:PM_depths}
    }
\end{figure}

To verify that the recovered PSF is optimal, we use the optical configuration in Fig.~\ref{sec:intro} with the optimized PSF to predict the depth at which an image with the real microscope was taken. We compare each image in the full depth range against a stack generated through our estimated model and plotted which depth yields the smallest error. As shown at the top of Fig.~\ref{fig:PSF_depths} the recovered PSF achieves the highest accuracy, with a mean NMSE error of $2.8\mu m$, against the initial PSF with error of $19.21\mu m$. Note how the ideal PSF is not robust against the depth symmetry, by predicting in some cases the wrong direction of defocus.

\subsection{Phase-mask distortion pattern recovery} \label{sec:sub:PM_estimation}
In this section we aim to recover the diffraction pattern created by the lack of 100\% pixel fill factor in a SLM \cite{Zhang2009,Liang2012} placed at the Fourier plane of a 4-f system (see supplementary material for an image of our setup). The zero-order diffraction lobe hits exactly the center of the output image, thus hampering the usability of an SLM. 

\subsubsection{Recovery on synthetic data} 
Consistently with sec.~\ref{sec:sub:PSF_estimation}, a test on synthetic data was first performed, where a PM was selected randomly between a cubic mask due to its simplicity and stable behavior in the Fourier domain \cite{Zhao2010} and a circular mask with a circular gradient towards the center, simulating the zero diffraction mode produced by a real SLM. Then a random scaling and translation where performed. A total of 500 examples where run, for which the results are presented in Table~\ref{table:synthResults}. A video of the optimization process can be found in the attached material.

\subsubsection{Recovering the phase-mask distortion pattern}
We use the recovered PSF from sec.~\ref{sec:sub:PSF_estimation} in a 4-f system, and propagate the WF until C2.
As discussed in sec.~\ref{sec:sub:PM_estimation}, the diffraction pattern created by the SLM distorts the PSF in the frequency domain and by recovering this pattern the distortion can be corrected. In this experiment we show that by using our synthesis model with the correction pattern displayed at the SLM an accurate depth of a stack of defocused images can be inferred (see Fig.~\ref{fig:PM_depths} bottom). Dealing a mean NMSE error of $0.67\mu m$ against $5.70 \mu m$ without PM correction. The recovered phase range matches the description of the vendor ($0$ to $5\pi$ max shift).

\section{Conclusions}
We have introduced a novel learning-based method to build a differentiable computational model of a real microscope. Experiments with both synthetic and real experiments, demonstrated that the proposed method allows to recover latent parameters of an optical setup (e.g., a PSF or a PM) with high accuracy.
We encourage the reader to explore the code repository\cite{WaveBlocks} for working examples and constant updates.
\bibliographystyle{IEEEbib}

\end{document}